\documentclass[11pt,a4paper]{article}
\usepackage{amsfonts}

\usepackage{amssymb}
\usepackage{graphicx}
\usepackage{amsmath}
\usepackage{makeidx}
\usepackage{indentfirst}
\usepackage[T1]{fontenc}
\usepackage[utf8]{inputenc}
\usepackage{authblk}

\newcounter{resultnum}[section]
\newcounter{conclusionnum}[section]
\newcounter{conditionnum}[section]
\newcounter{conjecturenum}[section]
\newcounter{examplenum}[section]
\newcounter{exercisenum}[section]
\newcounter{lemmanum}[section]
\newcounter{notationnum}[section]
\newcounter{theoremnum}[section]
\newcounter{definitionnum}[section]
\newcounter{corollarynum}[section]
\newcounter{remarknum}[section]
\newcounter{propositionnum}[section]
\newcounter{acknowledgementnum}[section]
\newcounter{algorithmnum}[section]
\newcounter{axiomnum}[section]
\newcounter{casenum}[section]
\newcounter{claimnum}[section]
\newcounter{summarynum}[section]
\newcounter{problemnum}[section]
\begin{document}

\title{Generic Off--Diagonal Solutions and\\
Solitonic Hierarchies in (Modified) Gravity}
\date{March 4, 2015}
\author{{\large \textbf{Sergiu I. Vacaru}}\thanks{%
sergiu.vacaru@cern.ch;\ sergiu.vacaru@uaic.ro} \\
%EndAName
{\qquad } \\
{\small Theory Division, CERN, CH-1211, Geneva 23, Switzerland} \thanks{%
associated visiting research}\\
{\small and } \\
{\small Rector's Department, Alexandru Ioan Cuza University}, \\
{\small Alexandru Lapu\c sneanu street, nr. 14, UAIC -- Corpus R, office 323}%
;\\
{\small Ia\c si,\ Romania, 700057} }
\maketitle

\begin{abstract}
There are summarized our recent results on encoding exact solutions of field
equations in Einstein and modified gravity theories into solitonic
hierarchies derived for nonholonomic curve flows with associated
bi--Hamilton structure. We argue that there is a canonical distinguished
connection for which the fundamental geometric/ physical equations decouple
in general form. This allows us to construct very general classes of generic
off--diagonal solutions determined by corresponding types of generating and
integration functions depending on all (spacetime) coordinates. If the
integral varieties are constrained to zero torsion configurations, we can
extract solutions for the general relativity theory. We conclude that the
geometric and physical data for various classes of effective/modified
Einstein spaces can be encoded into multi--component versions of the
sine--Gordon, or modified Korteweg -- de Vries equations.

\vskip0.1cm


\vskip3pt

\end{abstract}

\renewcommand\Authands{and }

%\title{Generic Off--Diagonal Solutions and Solitonic Hierarchies in (Modified) Gravity}

\vskip5pt

%\bigskip\bodymatter

\textbf{Preliminaries:}\ Let us consider a pseudo--Riemannian manifold $V$,%
\newline
$\dim V=n+m,$ ($n,m\geq 2$), endowed with a conventional nonholonomic
(equivalently, non--integrable, or anholonomic) horizontal (h) and vertical
(v) splitting of tangent bundle $TV$ via a Whitney sum $\mathbf{N}:\
TV=hTV\oplus vTV$. This defines a nonlinear connection, N--connection,
structure with coefficients $\mathbf{N}=\{N_{i}^{a}(u)\},$ when $\mathbf{N}%
=N_{i}^{a}(x,y)dx^{i}\otimes \partial /\partial y^{a}$ for local coordinates
$u=(x,y),$ or $u^{\alpha }=(x^{i},y^{a}),$ and indices $i,j,...=1,2,...n$
and $a,b,...=n+1,n+2,...,n+m.$ Boldface symbols will be used in order to
emphasize that the geometric objects are adapted to a N--splitting.

In N-adapted form, we can work equivalently with three "physically
important" linear connections completely defined by a metric structure $%
\mathbf{g}$: {\small
\begin{equation*}
\mathbf{g} \rightarrow \left\{
\begin{array}{ccc}
\mathbf{\nabla :} & \mathbf{\nabla g}=0;\ ^{\nabla }\mathcal{T}^{\alpha }=0,
& \mbox{ the Levi--Civita connection }; \\
\widehat{\mathbf{D}}: & \widehat{\mathbf{D}}\mathbf{g}=0;\ h\widehat{%
\mathcal{T}}^{\alpha }=0,v\widehat{\mathcal{T}}^{\alpha }=0, &
\mbox{ the
canonical d--connection }; \\
\grave{\mathbf{D}}: & \grave{\mathbf{D}}\mathbf{g}=0;\ h\grave{\mathcal{T}}%
^{\alpha }=0,v\grave{\mathcal{T}}^{\alpha }=0, & \mbox{ constant curvature
N--coefficients}.%
\end{array}%
\right.
\end{equation*}%
} The total torsions $\widehat{\mathcal{T}}$ and $\grave{\mathcal{T}}$ are
completely determined by $\mathbf{g}\sim \grave{\mathbf{g}}$ but for
different nonholonomic distributions $\mathbf{N}$ and/or $\grave{\mathbf{N}}$%
, which are equivalent up to frame transforms. There are unique distortion
relations $\widehat{\mathbf{D}}=\nabla + \widehat{\mathbf{Z}}$ and  $\grave{%
\mathbf{D}}=\nabla + \grave{\mathbf{Z}}$, where all connections and
distortion tensors $\widehat{\mathbf{Z}}$ and $\grave{\mathbf{Z}}$, and
related geometric and physical models, are completely defined by equivalent
geometric data $(\mathbf{g},\nabla)$ [standard general relativity, GR,
approaches], $(\mathbf{g},\mathbf{N},\widehat{\mathbf{D}})$ [convenient for
constructing off--diagonal solutions] and/or $(\grave{\mathbf{g}},\grave{%
\mathbf{N}},\grave{\mathbf{D}})$ [for encoding into solitonic hierarchies].
\newline
\vskip5pt \textbf{Decoupling of (Modified) Einstein eqs \& Off--Diagonal
Solutons:}\newline
The metrics are parameterized (we can put necessary "hat/grave" symbols)
\begin{eqnarray}
\mathbf{g} &=&h{g}+v{g}={g}_{ij}dx^{i}\otimes dx^{j}+{h}_{ab}\mathbf{e}%
^{a}\otimes \mathbf{e}^{b},\ {h}_{ab}={g}_{n+a\ n+b},  \label{lfsm} \\
\mathbf{e}_{\alpha } &=&(\mathbf{e}_{i}=\partial _{i}-{N}_{i}^{a}\partial
_{a},e_{a}=\partial _{a}),\ \mathbf{e}^{\alpha }=(e^{i}=dx^{i},\mathbf{e}%
^{a}=dy^{a}+{N}_{i}^{a}dx^{i}).  \notag
\end{eqnarray}

An Einstein manifold is defined by a solution $\mathbf{g}$ of the vacuum
gravitational field eqs with cosmological constant $\lambda $, $Ric=\lambda
\mathbf{g}$, where $Ric$ is the Ricci tensor of $\nabla$. These eqs can be
re--written equivalently in the form
\begin{equation}
\widehat{\mathbf{R}}ic = \lambda \mathbf{g}\mbox{ and } \mbox{(nonholonomic
constraints)\ } \ \widehat{\mathbf{Z}} = 0,\   \label{emdc}
\end{equation}%
for respective Ricci tensor, and/or (via nonholonomic frame deformations) in
constant curvature matrix coefficient form, $\grave{\mathbf{R}}ic =\lambda
\mathbf{g}$ and $\grave{\mathbf{Z}} = 0$. For $\widehat{\mathbf{Z}} = 0$,
the eqs (\ref{emdc}) define modified Einstein spaces with nonholonomically
induced torsion.

For simplicity, we consider solutions (\ref{lfsm}) for $n=2$, with Killing
symmetry on $\partial /\partial y^{4}$ (see ref. \cite%
{vexactsol,vexactsol1,vexactsol2,vexactsol3} for more general non--Killing
configurations depending on all spacetime coordinates) and parameterizations
$g_{ij}=diag[g_{i}=\epsilon _{i}e^{\psi }(x^{i})]$ and $%
h_{ab}=diag[h_{a}(x^{k},y^{3})];\ i,j,...=1,2;\ a,b,...=3,4;\epsilon _{i}=\pm 1
$ depending on chosen signature;\ $\mathbf{N}_{i}^{3}=w_{i}(x^{k},y^{3}),$ $%
\mathbf{N}_{i}^{4}=n_{i}(x^{k},y^{3})$; and use $a^{\bullet }=\partial
_{1}a,b^{\prime }=\partial _{2}b,h^{\ast }=\partial _{3}h$. Introducing $%
\alpha _{i}=h_{4}^{\ast }\partial _{i}\phi ,\beta =h_{4}^{\ast }\ \phi
^{\ast },\gamma =\left( \ln |h_{4}|^{3/2}/|h_{3}|\right) ^{\ast }$, we write
(\ref{emdc}) as
\begin{equation}
\epsilon _{1}\psi ^{\bullet \bullet }+\epsilon _{2}\psi ^{\prime \prime
}=2~\lambda ,\ \phi ^{\ast }h_{4}^{\ast }=2h_{3}h_{4}\lambda ,\ \beta
w_{i}-\alpha _{i}=0,\ n_{i}^{\ast \ast }+\gamma n_{i}^{\ast }=0,\
\label{eq4}
\end{equation}%
when the torsionless (Levi--Civita, LC) conditions $\widehat{\mathbf{Z}}=0$
are
\begin{eqnarray*}
&&w_{i}^{\ast }=(\partial _{i}-w_{i}\partial _{3})\ln \sqrt{|h_{3}|}%
,(\partial _{i}-w_{i}\partial _{3})\ln \sqrt{|h_{4}|}=0, \\
&&\partial _{i}w_{j}=\partial _{j}w_{i},n_{i}^{\ast }=0,\partial
_{i}n_{j}=\partial _{j}n_{i}.
\end{eqnarray*}

The system of PDEs (\ref{eq4}) can be integrated in very general forms. For
instance, the first 2--d Laplace/ d'Alamber equation can be solved for any
given $\lambda .$ Redefining the coordinates and $\Phi :=e^{{\phi }}$
(considered as a generating function) and introducing $\epsilon _{3}\epsilon
_{4}$ in $\lambda $, we express the solutions in functional form,  $%
h_{3}[\Phi ]=(\Phi ^{\ast })^{2}/\lambda \Phi ^{2},\ h_{4}[\Phi ]=\Phi
^{2}/4\lambda $. Solving respective algebraic equations, we compute  $%
w_{i}=\partial _{i}\phi /\phi ^{\ast }=\partial _{i}\Phi /\Phi ^{\ast }$.
Finally, integrating two times on $y^{3}$, we find  $n_{b}=\ _{1}n_{b}+\
_{2}n_{b}\int dy^{3}\ h_{3}/(\sqrt{|h_{4}|})^{3}$, where $\
_{1}n_{b}(x^{i}),\ _{2}n_{b}(x^{i})$ are integration functions.

The equations for nonholonomic for the LC--conditions can be solved in
explicit form for certain restricted classes of generating and integration
functions. For instance, taking $\tilde{\Phi}=\tilde{\Phi}(\ln \sqrt{|h_{3}|}%
)$ for a functional dependence $h_{3}[\tilde{\Phi}[\check{\Phi}]],$ when $%
(\partial _{i}\check{\Phi})^{\ast }=\partial _{i}(\check{\Phi}^{\ast }),$ we
can construct metrics of type {%\small
\begin{equation*}
ds^{2}=e^{\psi (x^{k})}\epsilon _{i}(dx^{i})^{2}+\epsilon _{3}\frac{(\check{%
\Phi}^{\ast })^{2}}{\lambda \check{\Phi}^{2}}[dy^{3}+(\partial _{i}%
\widetilde{A}[\check{\Phi}])dx^{i}]^{2}+\epsilon _{4}\frac{\check{\Phi}^{2}}{%
4|\lambda |}[dy^{4}+(\partial _{i}n)dx^{i}]^{2},  \label{qelgen}
\end{equation*}%
}for certain classes of integration functions $\ \ _{2}n_{i}=0$ and $\
_{1}n_{i}=\partial _{i}n$ with a function $n=n(x^{k}).$ We should take $%
w_{i}=\check{w}_{i}=\partial _{i}\check{\Phi}/\check{\Phi}^{\ast }=\partial
_{i}\widetilde{A}$, with a nontrivial function $\widetilde{A}(x^{k},y^{3})$
depending functionally on generating function $\check{\Phi}$, and satisfy $%
\partial _{i}w_{j}=\partial _{j}w_{i}$.

\vskip5pt \textbf{Generating Solitonic Hierarchies:}\ A non--stretching
curve $\gamma (\tau ,\mathbf{l})$ on a (modified) Einstein manifold $\mathbf{%
V,}$ where $\tau $ is a real parameter and $\mathbf{l}$ is the arclength of
the curve on a nonholonomic $\mathbf{V,}$ is defined with such evolution
d--vector $\mathbf{Y}=\gamma _{\tau }$ and tangent d--vector $\mathbf{X}%
=\gamma _{\mathbf{l}}$ that $\mathbf{g(X,X)=}1\mathbf{.}$ Such a curve $%
\gamma (\tau ,\mathbf{l})$ swept out a two--dimensional surface in $%
T_{\gamma (\tau ,\mathbf{l})}\mathbf{V}\subset T\mathbf{V.}$ We consider a
coframe $\mathbf{e}\in T_{\gamma }^{\ast }\mathbf{V}_{\mathbf{N}}\otimes (h%
\mathfrak{p\oplus }v\mathfrak{p}),$ which is a N--adapted $\left( SO(n)%
\mathfrak{\oplus }SO(m)\right) $--parallel basis along $\gamma $, see
details in ref. \cite{vsolithierarch,vsolithierarch1,vsolithierarch2}. The
associated linear connection 1--form is $\mathbf{\Gamma }\in T_{\gamma
}^{\ast }\mathbf{V}_{\mathbf{N}}\otimes (\mathfrak{so}(n)\mathfrak{\oplus so}%
(m))$ (we can put "hat" or "grave" label on such geometric objects). We
parameterize $\ \mathbf{e}_{\mathbf{X}}=\mathbf{e}_{h\mathbf{X}}+\mathbf{e}%
_{v\mathbf{X}},$ where (for $(1,\overrightarrow{0})\in \mathbb{R}^{n},%
\overrightarrow{0}\in \mathbb{R}^{n-1}$ and $(1,\overleftarrow{0})\in
\mathbb{R}^{m},\overleftarrow{0}\in \mathbb{R}^{m-1}),$ for {\small
\begin{equation*}
\mathbf{e}_{h\mathbf{X}}=\gamma _{h\mathbf{X}}\rfloor h\mathbf{e=}\left[
\begin{array}{cc}
0 & (1,\overrightarrow{0}) \\
-(1,\overrightarrow{0})^{T} & h\mathbf{0}%
\end{array}%
\right] ,\mathbf{e}_{v\mathbf{X}}=\gamma _{v\mathbf{X}}\rfloor v\mathbf{e=}%
\left[
\begin{array}{cc}
0 & (1,\overleftarrow{0}) \\
-(1,\overleftarrow{0})^{T} & v\mathbf{0}%
\end{array}%
\right] .
\end{equation*}%
} For a $n+m$ splitting, ${\mathbf{\Gamma }}=\left[ \mathbf{\Gamma }_{h%
\mathbf{X}},\mathbf{\Gamma }_{v\mathbf{X}}\right] $, with
\begin{equation*}
{\mathbf{\Gamma }}_{h\mathbf{X}}\mathbf{=}\gamma _{h\mathbf{X}}\rfloor
\mathbf{L=}\left[
\begin{array}{cc}
0 & (0,\overrightarrow{0}) \\
-(0,\overrightarrow{0})^{T} & \mathbf{L}%
\end{array}%
\right] \in \mathfrak{so}(n+1),
\end{equation*}%
where $\mathbf{L=}\left[
\begin{array}{cc}
0 & \overrightarrow{v} \\
-\overrightarrow{v}^{T} & h\mathbf{0}%
\end{array}%
\right] \in \mathfrak{so}(n),~\overrightarrow{v}\in \mathbb{R}^{n-1},~h%
\mathbf{0\in }\mathfrak{so}(n-1),$ and
\begin{equation*}
{\mathbf{\Gamma }}_{v\mathbf{X}}\mathbf{=}\gamma _{v\mathbf{X}}\rfloor
\mathbf{C=}\left[
\begin{array}{cc}
0 & (0,\overleftarrow{0}) \\
-(0,\overleftarrow{0})^{T} & \mathbf{C}%
\end{array}%
\right] \in \mathfrak{so}(m+1),
\end{equation*}%
where $\mathbf{C=}\left[
\begin{array}{cc}
0 & \overleftarrow{v} \\
-\overleftarrow{v}^{T} & v\mathbf{0}%
\end{array}%
\right] \in \mathfrak{so}(m),~\overleftarrow{v}\in \mathbb{R}^{m-1},~v%
\mathbf{0\in }\mathfrak{so}(m-1).$

The canonical d--connection induces matrices decomposed with respect to the
flow direction:\ in the h--direction,
\begin{equation*}
\mathbf{e}_{h\mathbf{Y}}=\gamma _{\tau }\rfloor h\mathbf{e=}\left[
\begin{array}{cc}
0 & \left( h\mathbf{e}_{\parallel },h\overrightarrow{\mathbf{e}}_{\perp
}\right) \\
-\left( h\mathbf{e}_{\parallel },h\overrightarrow{\mathbf{e}}_{\perp
}\right) ^{T} & h\mathbf{0}%
\end{array}%
\right],
\end{equation*}
when $\mathbf{e}_{h\mathbf{Y}}\in h\mathfrak{p,}\left( h\mathbf{e}%
_{\parallel },h\overrightarrow{\mathbf{e}}_{\perp }\right) \in \mathbb{R}%
^{n} $ and $h\overrightarrow{\mathbf{e}}_{\perp }\in \mathbb{R}^{n-1},$ and
\begin{equation*}
{\mathbf{\Gamma }}_{h\mathbf{Y}}\mathbf{=}\gamma _{h\mathbf{Y}}\rfloor
\mathbf{L=}\left[
\begin{array}{cc}
0 & (0,\overrightarrow{0}) \\
-(0,\overrightarrow{0})^{T} & h\mathbf{\varpi }_{\tau }%
\end{array}%
\right] \in \mathfrak{so}(n+1),
\end{equation*}
where $\ h\mathbf{\varpi }_{\tau }\mathbf{=}\left[
\begin{array}{cc}
0 & \overrightarrow{\varpi } \\
-\overrightarrow{\varpi }^{T} & h\mathbf{\Theta }%
\end{array}%
\right] \in \mathfrak{so}(n),~\overrightarrow{\varpi }\in \mathbb{R}^{n-1},~h%
\mathbf{\Theta \in }\mathfrak{so}(n-1).$

Similar parameterizations can be performed in the v--direction,
\begin{equation*}
\mathbf{e}_{v\mathbf{Y}}=\gamma _{\tau }\rfloor v\mathbf{e=}\left[
\begin{array}{cc}
0 & \left( v\mathbf{e}_{\parallel },v\overleftarrow{\mathbf{e}}_{\perp
}\right) \\
-\left( v\mathbf{e}_{\parallel },v\overleftarrow{\mathbf{e}}_{\perp }\right)
^{T} & v\mathbf{0}%
\end{array}%
\right],
\end{equation*}
when $\mathbf{e}_{v\mathbf{Y}}\in v\mathfrak{p,}\left( v\mathbf{e}%
_{\parallel },v\overleftarrow{\mathbf{e}}_{\perp }\right) \in \mathbb{R}^{m}$
and $v\overleftarrow{\mathbf{e}}_{\perp }\in \mathbb{R}^{m-1},$ and
\begin{equation*}
{\mathbf{\Gamma }}_{v\mathbf{Y}}\mathbf{=}\gamma _{v\mathbf{Y}}\rfloor
\mathbf{C=}\left[
\begin{array}{cc}
0 & (0,\overleftarrow{0}) \\
-(0,\overleftarrow{0})^{T} & v\mathbf{\varpi }_{\tau }%
\end{array}%
\right] \in \mathfrak{so}(m+1),
\end{equation*}
where $v\mathbf{\varpi }_{\tau }\mathbf{=}\left[
\begin{array}{cc}
0 & \overleftarrow{\varpi } \\
-\overleftarrow{\varpi }^{T} & v\mathbf{\Theta }%
\end{array}%
\right] \in \mathfrak{so}(m),~\overleftarrow{\varpi }\in \mathbb{R}^{m-1},~v%
\mathbf{\Theta \in }\mathfrak{so}(m-1).$

\vskip5pt \textbf{Main Results: } For any solution for (modified) Einstein
manifolds, there is a hierarchy of N--adapted flows of curves ${\gamma }%
(\tau ,\mathbf{l})=h{\gamma }(\tau ,\mathbf{l})+v{\gamma }(\tau ,\mathbf{l})$
described by geometric nonholonomic map equations:

\begin{itemize}
\item The $0$ flows are convective (travelling wave) maps  ${\gamma }_{\tau
}={\gamma }_{\mathbf{l}}$  distinguished as $\left( h{\gamma }\right) _{\tau
}=\left( h{\gamma }\right) _{h\mathbf{X}}$ and $\left( v{\gamma }\right)
_{\tau }=\left( v{\gamma }\right) _{v\mathbf{X}}$.

\item There are +1 flows defined as non--stretching mKdV maps,
\begin{eqnarray*}
&&-\left( h{\gamma }\right) _{\tau } ={\mathbf{D}}_{h\mathbf{X}}^{2}\left( h{%
\gamma }\right) _{h\mathbf{X}}+\frac{3}{2} |{\mathbf{D}}_{h\mathbf{X}}\left(
h{\gamma }\right) _{h\mathbf{X}}|_{h\mathbf{g}}^{2}~\left( h{\gamma }\right)
_{h\mathbf{X}} \mbox{ and } \\
&&-\left( v{\gamma }\right) _{\tau } ={\mathbf{D}}_{v\mathbf{X}}^{2}\left( v{%
\gamma }\right) _{v\mathbf{X}}+\frac{3}{2} |{\mathbf{D}}_{v\mathbf{X}}\left(
v{\gamma }\right) _{v\mathbf{X}}|_{v\mathbf{g}}^{2}~\left( v{\gamma }\right)
_{v\mathbf{X}},
\end{eqnarray*}
and the +2,... flows as higher order analogs.

\item Finally, the -1 flows are defined by the kernels of the recursion
h--operator,
\begin{equation*}
h{\mathfrak{R}} ={\mathbf{D}}_{h\mathbf{X}}\left( {\mathbf{D}}_{h\mathbf{X}}+%
{\mathbf{D}}_{h\mathbf{X}}^{-1}\left( \overrightarrow{v}\cdot \right)
\overrightarrow{v}\right) +\overrightarrow{v}\rfloor {\mathbf{D}}_{h\mathbf{X%
}}^{-1}\left( \overrightarrow{v}\wedge {\mathbf{D}}_{h\mathbf{X}}\right),
\end{equation*}
and of the recursion v--operator,
\begin{equation*}
v{\mathfrak{R}} = {\mathbf{D}}_{v\mathbf{X}}\left({\mathbf{D}}_{v\mathbf{X}}+%
{\mathbf{D}}_{v\mathbf{X}}^{-1}\left( \overleftarrow{v}\cdot \right)
\overleftarrow{v}\right) +\overleftarrow{v}\rfloor {\mathbf{D}}_{v\mathbf{X}%
}^{-1}\left( \overleftarrow{v}\wedge {\mathbf{D}}_{v\mathbf{X}}\right),
\end{equation*}
inducing non--stretching maps  ${\mathbf{D}}_{h\mathbf{Y}}\left( h{\gamma }%
\right) _{h\mathbf{X}}=0$ and ${\mathbf{D}}_{v\mathbf{Y}}\left( v{\gamma }%
\right) _{v\mathbf{X}}=0$.
\end{itemize}

\vskip5pt \textbf{Acknowledgments:} Author's research is partially supported by IDEI,
PN-II-ID-PCE-2011-3-0256 and communicated at Parallel Section GT1 -- Exact
Solutions in Four and Higher Dimensions: Mathematical Aspects; Thirteen
Marcel Grossmann Meeting -- MG13, Stockholm University, Sweden; July 1-7,
2012. The approach with nonolonomic configurations and applications in
modified gravity was elaborated during a short term visit at CERN-TH in October-November 2013.


\begin{thebibliography}{9}
\bibitem{vexactsol} S. Vacaru, \textsl{IJGMMP} \textbf{8}, 9 (2011).\

\bibitem{vexactsol1} S. Vacaru, \textsl{Eur. Phys. J.} \textbf{C 73}, 2287
(2013).

\bibitem{vexactsol2} T. Gheorgiu, O. Vacaru and S. Vacaru, \ \textsl{Eur. Phys. J.} \textbf{C74}, 3152 (2014).

\bibitem{vexactsol3} T. Gheorgiu, O. Vacaru and S. Vacaru, \ \textsl{Class. Quant. Grav.}  \textbf{C32}, 065004 (2015).

\bibitem{vsolithierarch} S. Vacaru, \textsl{Acta Applicandae Mathematicae}
\textbf{110}, 73 (2010).\

\bibitem{vsolithierarch1} S. Vacaru, \textsl{Chaos, Solitons \& Fractals}
\textbf{45}, 1266 (2012).\

\bibitem{vsolithierarch2} S. Anco and S. Vacaru, J. Geom. Phys. \textbf{59},
79 (2009).
\end{thebibliography}
\end{document}